\documentclass[%
 reprint,
showkeys,
 amsmath,amssymb,
 aps,
]{revtex4-1}

\usepackage{graphicx}
\usepackage{dcolumn}
\usepackage{bm}

\newcommand{\degree}{\({}^\circ\)C}

\newcommand{\etal}{{\itshape et al.}}

\begin{document}

\preprint{APS/123-QED}

\title{
Dynamic shear responses of polymer-polymer interfaces
}%

\author{Yasuya Nakayama}
 \email{nakayama@chem-eng.kyushu-u.ac.jp}
\author{Kiyoyasu Kataoka}
\author{Toshihisa Kajiwara}
\affiliation{%
Department of Chemical Engineering,
Kyushu University,
Nishi-ku,
Fukuoka 819-0395,
Japan
}%


\date{May 5, 2012}

\begin{abstract}
In multi-component soft matter, interface properties often play a key role in determining the properties of the overall system. The identification of the internal dynamic structures in non-equilibrium situations requires the interface rheology to be characterized. We have developed a method to quantify the rheological contribution of soft interfaces and evaluate the dynamic modulus of the interface.
This method reveals that the dynamic shear responses of interfaces in bilayer systems comprising polypropylene and three different polyethylenes can be classified as having hardening and softening effects on the overall system:
a interface between linear long polymers becomes more elastic than the
component polymers, 
while large polydispersity or long-chain-branching of one component make
the interface more viscous.
We find that the chain lengths and architectures of the component polymers, rather than equilibrium immiscibility, play an essential role in determining the interface rheological properties.
\end{abstract}

\maketitle


\section{Introduction}
The rheological properties of emulsions or blends of immiscible materials have been the subject of much research.
The physical properties of such multi-component complexes are not only a simple average of the properties of the components but also depend on the morphological structure and interface properties.
Many studies have addressed the relationship between morphology and
rheological properties.\cite{Sperling1997Polymeric,Robeson2007Polymer}
In contrast, the rheological properties of an interface are still not fully understood.

Regarding the extrusion instability of a polymer melt, slippage at a
polymer--wall interface has received much
attention.\cite{Brochard1992Sheardependent,Munstedt2000Stick,Denn2001EXTRUSION,Migler1993Slip,Park2008Wall}
In contrast, slippage at a polymer--polymer interface was introduced to
explain the anomalously low viscosity in immiscible polymer
blends.\cite{ChenChongLin1979Mathematical,LyngaaeJorgensen1988Influence}
Considering shear flow parallel to the flat interface between two polymers, the viscosity of this bilayer system under the stick boundary condition, \(\eta^{\text{st}}\), becomes the harmonic mean of the viscosities of the components, \(\eta_{1}\) and \(\eta_{2}\):
\begin{align}
 \frac{1}{\eta^{\text{st}}}&=\sum_{\alpha=1,2}\frac{\phi_{\alpha}}{\eta_{\alpha}},
\end{align}
where \(\phi_{\alpha}\) is the volume fraction of the polymer
\(\alpha\).
Lin defined the ratio of the stick viscosity to the measured viscosity,
\(\eta\), to characterize the slip:
\(\beta=\eta^{\text{st}}/\eta\).\cite{ChenChongLin1979Mathematical}
The physical origin of the partial slip of \(\beta>1\) was attributed to the viscosity of the interfacial layer, \(\eta_{I}\), by Lyngaae-J{\o}rgensen \etal\cite{LyngaaeJorgensen1988Influence} as
\begin{align}
 \frac{1}{\eta}&=\sum_{\alpha=1,2}\frac{\phi_{\alpha}}{\eta_{\alpha}}
+\frac{\phi_{I}}{\eta_{I}},
\label{eq:viscosity_interface}
\end{align}
where \(\phi_{I}\) is the volume fraction of the interfacial layer.
Rewriting Eq. (\ref{eq:viscosity_interface}) with \(\beta\) under the thin-interface limit of \(\phi_{I}\to 0\) gives the ratio of the bilayer viscosity to the interface viscosity,
\(1-1/\beta=\eta\phi_{I}/\eta_{I}\).
Let the interface thickness be \(h_{I}\); the slip velocity of the interface, \(\Delta V_{I}=h_{I}\dot{\gamma}_{I}\), can be defined to be finite even under the thin-interface limit of \(h_{I}\to 0\).
In a bilayer system with the thickness \(h_{W}\), let the shear stress be \(\tau\) and the apparent shear rate of the bilayer be \(\dot{\gamma}_{W}\). We have the relation \(\eta\phi_{I}/\eta_{I}=\dot{\gamma}_{I}h_{I}\tau/(\dot{\gamma}_{W}h_{W}\tau)=\Delta V_{I}/\Delta V_{W}\), where \(\Delta V_{W}\) is the relative velocity between both surfaces of the bilayer.
Lam \etal~redefined the degree of slip as \(\varphi=1-1/\beta= \Delta
V_{I}/\Delta V_{W}\), which they called the energy dissipation factor in
relation to the contribution of the interface to the total energy
dissipation of the bilayer system.\cite{Lam2003Interfacial_properties}
Along with the energy dissipation factor, another slip index \(\phi_{I}\gamma_{I}/\gamma_{W}\) under oscillatory shear deformation parallel to the interface was used\cite{Jiang2005Rheological}, where \(\gamma_{I}\) and \(\gamma_{W}\) are the strain amplitudes of the interface layer and the entire bilayer, respectively.
The existence of the slippage at different polymer--polymer interfaces
have been reported by observing non-zero
\(\varphi\).\cite{Lam2003Interfacial_properties,Jiang2005Rheological,Lam2003Interfacial,Jiang2003Energy}

The slip velocity's onset and dependence on the shear stress have also
been studied.\cite{Migler2001Visualizing,Zhao2002Slip,lee09:_polym_polym_inter_slip_in_multil_films,Park2010Polymerpolymer}
These studies focused on the slip at rather high shear rate. 
The sigmoidal dependence of \(\Delta V_{I}\) on \(\tau\) and power-law regimes of \(\Delta V_{I}\propto \tau^{n}\) were observed in different polymer--polymer interfaces, and the exponent \(n\) depends on pairs.
The differences in exponents might be related to the miscibility, molecular weight distributions and entanglement structures of the components, but this matter still requires more systematic studies.
A polymer--polymer interface has a small but finite thickness.\cite{Jones1999Polymers}
Thus, different conformations from the bulk in a certain region around
an interface may induce different relaxations, {\itshape i.e.}, different viscoelastic properties.
From this viewpoint, the slip velocity under steady shear determines the viscous property of the interface.
For bulk polymers, the dynamic response reflects the internal structure of a material; therefore, rheological measurements are used to assess structural information.\cite{dealy06:_struc_and_rheol_of_molten_polym}
Thus, the dynamic response of a polymer--polymer interface can be used as a probe of the interface structure.

An attempt to measure the dynamic modulus of polymer--polymer interfaces was reported.
Song and Dai\cite{Song2010TwoParticle} applied a passive microrheology technique to an interface between polydimethylsiloxane and polyethylene glycol using Pickering emulsions at room temperature. 
In this technique, a tracer particle is confined to the interface region by physico-chemical absorption.

In this article, we developed a method to evaluate the dynamic modulus of a polymer--polymer interface in the linear-response regime.
To form an interface for study, bilayer samples of immiscible polymers were prepared.
From independent dynamic measurements of a bilayer and its component polymers, the dynamic modulus of an interface was evaluated.
This technique was applied to interfaces between polypropylene and different polyethylenes with different chain architectures and molecular weight distributions.
\section{Interfacial boundary layer and its dynamic modulus}

We consider a system with two immiscible polymers melts A and B having a macroscopically flat interface.
In equilibrium, the interface between the two immiscible polymers has a small but finite thickness of \(a_{I}\).\cite{Jones1999Polymers}

\begin{figure}[htbp]
\center
 \includegraphics[width=.7\hsize,angle=-90]{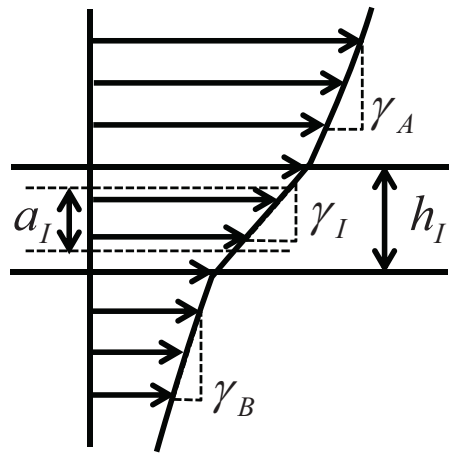}
\caption{Schematic of the displacement across an interfacial boundary layer.}
\label{fig:interfacial_boundary_layer}
\end{figure}

To discuss the dynamic response of the interface, we apply an oscillatory shear stress 
 \begin{align}
  \tau(\omega, \tau_{0})&=\tau_{0}e^{i\omega t},
 \end{align}
parallel to the interface on the polymer A and measure a response strain of the overall bilayer system, \(\gamma_{W}\).
In this measurement, the applied stress on the polymer A is transmitted to the polymer B through the interface.
It is assumed that a region around the interface exists whose response strain, \(\gamma_{I}\), to the applied shear stress is different from those of the A and B bulk layers, \(\gamma_{A}\) and \(\gamma_{B}\).
We call this region the {\itshape interfacial boundary layer}~(Fig. \ref{fig:interfacial_boundary_layer}).
The thickness of the interfacial boundary layer \(h_{I}\), which is defined under non-equilibrium steady state, may be different from the equilibrium thickness \(a_{I}\), which is determined thermodynamically by the free energy of the entire bilayer system.
Similarly, a layer deforming with the strain response of the bulk
polymer A(B), \(\gamma_{A}\)(\(\gamma_{B}\)), is called the layer A(B).
The thickness of the layer A(B) is denoted as \(h_{A}\)(\(h_{B}\)).
Therefore, the thickness of the overall system is \(h_{W}=h_{A}+h_{I}+h_{B}\).

We then define the thickness fraction as
\begin{align}
 c_{\alpha}&=\frac{h_{\alpha}}{h_{W}}, ~~(\alpha=A,B,I)
\end{align}
where the relation \(1=\sum_{\alpha=A,B,I}c_{\alpha}\) holds.
With these thickness fractions, the apparent deformation \(\gamma_{W}\) of the overall system is expressed as
\begin{align}
 \gamma_{W}&=
 \sum_{\alpha=A,B,I}\gamma_{\alpha}c_{\alpha},
 \label{eq:strain_total_system}
\end{align}
We note that, although the thickness of the interfacial boundary layer is supposed to be much smaller than those of the bulk layers~(\(c_{I}\ll
c_{A}, c_{B}\)), the deformation of the interfacial boundary layer can be comparable to those of the bulk layers.
Suppose that \(\gamma_{I}\approx \gamma_{A}, \gamma_{B}\).  The contribution of the interfacial boundary layer to the total strain is much smaller than those of the bulk layers due to the small thickness of the interfacial boundary layer, {\itshape i.e.},  \(c_{I}\gamma_{I} \ll
c_{A}\gamma_{A}, c_{B}\gamma_{B}\) and is therefore barely detectable macroscopically.
In this case, the interface would be referred to as macroscopically stick.
In contrast, suppose that \(c_{I}\gamma_{I}\gtrsim c_{A}\gamma_{A}, c_{B}\gamma_{B}\), where the deformation of the interfacial boundary layer is comparable to that of the bulk layers.
In this case, the interface would be referred to as macroscopically non-stick or slip.

We are ready to consider the dynamic modulus of each layer.
The time-dependent strain of the \(\alpha\)th layer is written as
\(\gamma_{\alpha}=\gamma_{\alpha,0}e^{i\omega t}\) with a 
time-independent amplitude \(\gamma_{\alpha,0}\).
The complex modulus of the \(\alpha\)th layer in the linear response
regime, \(G_{\alpha}^{*}(\omega)\), is defined as
\begin{align}
 \tau&=G_{\alpha}^{*}\gamma_{\alpha,0}e^{i\omega t},
 \label{eq:complex_modulus_alpha}
\\
 G_{\alpha}^{*}&=
 \left|G_{\alpha}^{*}\right|e^{i\delta_{\alpha}}
 =
 G_{\alpha}^{'}+i G_{\alpha}^{''},
\end{align}
where \(G'_{\alpha}\), \(G''_{\alpha}\), and \(\delta_{\alpha}\) are the storage modulus, loss modulus, and phase lag of the \(\alpha\)th layer, respectively.
Combining Eqs. (\ref{eq:strain_total_system}) and
(\ref{eq:complex_modulus_alpha})
yields
\begin{align}
 \frac{G_{I}^{*}}{c_{I}}&=\left[
 \frac{1}{G_{W}^{*}}-\sum_{\alpha=A,B}
 \frac{c_{\alpha}}{G_{\alpha}^{*}}
 \right]^{-1},
 \label{eq:interface_modulus0}
\end{align}
which is the dynamic modulus of the interfacial boundary layer up to the interface thickness fraction.
Let
\begin{align}
 D_{s} -iD_{L} &=
 \frac{1}{G_{W}^{*}}-\sum_{\alpha=A,B}
 \frac{c_{\alpha}}{G_{\alpha}^{*}},
 \label{eq:def_ds_dl}
\end{align}
then the amplitude and the phase lag of the interface dynamic modulus are expressed as
\begin{align}
 \frac{\left|
 G_{I}^{*}
 \right|}{c_{I}}&=\frac{1}{\sqrt{
 D_{S}^{2}
 +
 D_{L}^{2}
 }},
 \label{eq:interface_modulus_ds_dl}
 \\
 \tan\delta_{I} &= \frac{D_{L}}{D_{S}}.
 \label{eq:interface_tandelta_ds_dl}
\end{align}
\(D_{S}\) and \(D_{L}\) are determined by \(G_{W}^{*}\), \(G_{A}^{*}\) ,
\(G_{B}^{*}\), \(c_{A}\) and \(c_{B}\). 
Henceforth, Eqs. (\ref{eq:interface_modulus_ds_dl}) and
(\ref{eq:interface_tandelta_ds_dl}) are used to determine the dynamic
modulus of the interfacial boundary layer up to \(c_{I}\).
Note that the phase lag of the interfacial boundary layer can be determined independently of \(c_{I}\).

\section{Assessment of the interfacial contribution to the overall bilayer system}
To obtain the dynamic modulus of the interfacial boundary layer, the thickness fractions, \(c_{A}\) and \(c_{B}\) , in addition to the moduli \(G_{W}^{*}\), \(G_{A}^{*}\), and \(G_{B}^{*}\), are required.
In this section, we introduce a method of assessing the presence and extent of the
rheological contribution of the interface to the overall bilayer system solely
from the moduli \(G_{W}^{*}\), \(G_{A}^{*}\), and \(G_{B}^{*}\).

\begin{figure}[htbp]
\center
\includegraphics[width=.7\hsize]{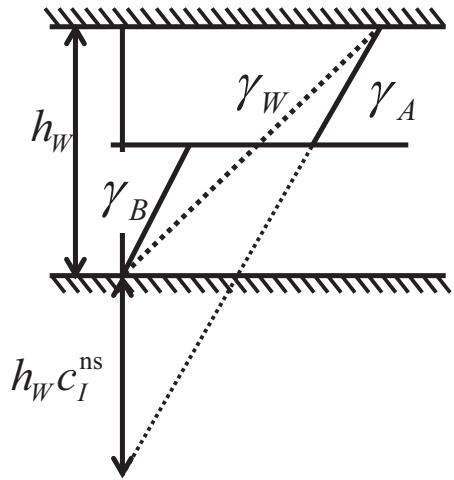}
\caption{Schematic illustration of displacement for a negative non-stick
 degree of interface, 
\(c_{I}^{\text{ns}}>0\).  To explain the apparent strain for the
bilayer system \(\gamma_{W}\) from the strains \(\gamma_{A}\) and
\(\gamma_{B}\), large strain at the thin interface is implied.  }
\label{fig:cins_softening}
\end{figure}

\begin{figure}[htbp]
\center
\includegraphics[width=.7\hsize,angle=0]{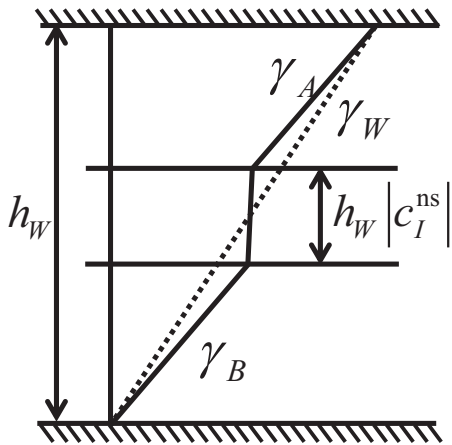}
\caption{Schematic illustration of displacement for a positive non-stick
 degree of interface, 
\(c_{I}^{\text{ns}}<0\).  
To explain the
 apparent strain of the bilayer system \(\gamma_{W}\) from the strains
 \(\gamma_{A}\) and \(\gamma_{B}\), a finite interfacial boundary layer
 with a small strain is implied.  }
\label{fig:cins_hardening}
\end{figure}

Assume that no interfacial boundary layer exists under shear such that the interface exhibits a completely stick response.
In this case, the modulus of the overall system can be completely
determined by those of the bulk layers of the polymers A and B.
Equation~(\ref{eq:interface_modulus0}) is reduced to 
\begin{align}
 \frac{e^{-i\delta_{W}}}{\left|G_{W}^{*}\right|}
&=
\sum_{k=A,B}
 c_{k}\frac{e^{-i\delta_{k}}}{\left|G_{k}^{*}\right|}.
 \label{eq:interface_modulus_stick}
\end{align}
We note that this relation~(\ref{eq:interface_modulus_stick}) can be
regarded as a linear simultaneous equation with two unknowns \(c_{A}\)
and \(c_{B}\) for given \(G_{W}^{*}\), \(G_{A}^{*}\), and \(G_{B}^{*}\).
Let \(c_{A}^{\text{st}}\) and \(c_{B}^{\text{st}}\) be the solutions of
Eq. (\ref{eq:interface_modulus_stick}).
If the assumption of the stick interface, namely no interface
contribution, is valid, 
the relation \(c_{A}^{\text{st}} + c_{B}^{\text{st}}=1\) is expected.
On the contrary, if an interfacial contribution exists, this effect is
reflected in \(c_{A}^{\text{st}} + c_{B}^{\text{st}}\neq 1\).
From this observation, we define a measure of the interface contribution by
\(c_{I}^{\text{ns}}=c_{A}^{\text{st}} + c_{B}^{\text{st}}-1\), 
which we call the {\itshape non-stick degree of interface}.
To understand the physical implication of \(c_{I}^{\text{ns}}\), we consider a model case of two immiscible polymers with the same modulus \(G_{A}^{*}=G_{B}^{*}\).
For simplicity, we assume that \(\delta_{W}=\delta_{A}\).
In this case, Eq. (\ref{eq:interface_modulus_stick}) reduces to 
\begin{align}
\frac{\left|G_{A}^{*}\right|}{\left|G_{W}^{*}\right|}
&=
c_{A}^{\text{st}} + c_{B}^{\text{st}} = 1+c_{I}^{ns}.
\label{eq:stick_modulus}
\end{align}
When the interface is stick,
\(\left|G_{W}^{*}\right|=\left|G_{A}^{*}\right|\) implies that
\(c_{I}^{\text{ns}}=0\).
When the interface is non-stick, both positive and negative \(c_{I}^{\text{ns}}\) are possible.
A positive \(c_{I}^{\text{ns}}\) implies that \(\left|G_{W}^{*}\right|<\left|G_{A}^{*}\right|\), which means that the bilayer is softened due to the existence of the interfacial boundary layer~(Fig. \ref{fig:cins_softening}).
In this case, the excess length of 
\(h_{W}c_{I}^{\text{ns}}\)
is a counterpart of the slip length, \(l=\Delta V_{I}/\dot{\gamma}\), in
slippage under steady shear.
\cite{gennes79:_viscom_flows_of_tangl_polym}
Conversely,  a negative \(c_{I}^{\text{ns}}\)  implies that
\(\left|G_{W}^{*}\right|>\left|G_{A}^{*}\right|\), which means that the
bilayer is hardened due to the existence of the interfacial boundary
layer~(Fig. \ref{fig:cins_hardening}).
In this case, 
to explain the apparent strain of the bilayer system,
 \(\gamma_{W}\), solely from the strains \(\gamma_{A}\) and \(\gamma_{B}\),
a finite interfacial boundary layer is required;
a layer with a finite thickness 
of \(h_{W}\left|c_{I}^{\text{ns}}\right|\) 
has zero strain.
For the interface modulus to be finite, a thickness of the interfacial
boundary layer 
\(h_{I}>h_{W}\left|c_{I}^{\text{ns}}\right|\) is implied.
In other words, \(h_{W}\left|c_{I}^{\text{ns}}\right|\) gives a lower
limit for the thickness of the interfacial boundary layer.
In general, a nonzero \(c_{I}^{\text{ns}}\) implies the existence of an interface
contribution to the bilayer modulus. 
Moreover, the different sign of \(c_{I}^{\text{ns}}\) reflects a different qualitative
contribution of the interface to the bilayer modulus.


\section{Experimental}

\subsection{Materials}
The polymers used in this study are listed in Table~\ref{tbl:samples}.
Three different polyethylenes were obtained from the Japan Polyethylene Corporation:
linear low-density polyethylene~(LLDPE)(
NOVATEC{\texttrademark}-LL(UJ960)),
high-density polyethylene~(HDPE)(
NOVATEC{\texttrademark}-HD(HJ362N)), 
and
low-density polyethylene~(LDPE)(
NOVATEC{\texttrademark}-LD(LF640MA)).
Polypropylene~(PP) was obtained from the Japan Polypropylene
Corporation~(NOVATEC{\texttrademark}-PP~(BC4BSW))
The molecular weight distributions 
of LLDPE, HDPE, and PP 
were
estimated from their melt dynamic moduli by an algorithm developed by
Mead.\cite{Mead1994Determination}
This method has been implemented in Rheometric Scientific's
Orchestrator software of TA instruments.
The Mead's algorithm is based on theory for a linear polymer and cannot
be used for a polymer with long-chain-branching.
The weight-averaged molecular weight,~\(M_{w}\), and polydispersity,
\(M_{w}/M_{n}\), of LDPE was supplied by the manufacturer.
The polymers were in pellet form.
The melting point, \(T_{m}\), and thermal decomposition temperature were
obtained through thermogravimetry and differential thermal analysis in
air.

\begin{table}[htbp]
\center
\caption{Characteristics of the polymer samples.
Molecular weight distributions estimated
 from the melt dynamic moduli by an algorithm developed by
 Mead.\cite{Mead1994Determination},
with the exception of LDPE, for which the manufacturer supplied the
information. 
}
\label{tbl:samples}
\begin{tabular}{ccccc}
\hline
polymer & \(M_{w}\)(kg/mol) & \(M_{w}/M_{n}\) & \(T_{m}\)(\degree) & \(T_{d}\)(\degree) \\
\hline
\hline
 polyethylene & & & & \\ 
 LLDPE
& 71.2
& 4.49
& 127 & 232 \\ 
 HDPE
& 70.4
& 13.3
& 132 & 220 \\ 
 LDPE
 & 65
&
 3.82
 & 114 & 229 \\ 
 polypropylene & & & & \\ 
 PP
& 155
& 9.85
 & 168 & 236 \\ 
\hline
\end{tabular}
\end{table}
\subsection{Sample Preparation}
Polymer pellets were dried for at least 24~h at 100\degree~in
a vacuum oven prior to use.
Each polymer was compression-molded at a temperature of approximately
60\degree~above each melting point in a hot press to form a plaque
and then cooled to room temperature.
The thickness of each polymer plaque was kept at approximately 1~mm
using a spacer frame made of stainless steel so that the plaque was suitable for the
rheological measurements described below.
Each polymer melt was pressed between two sheets of
polytetrafluoroethylene~(Naflon\texttrademark PTFE from NICHIAS
Corporation).
\subsection{Rheological measurements}

We focused on pairs of polypropylene and different
polyethylenes, namely PP/LLDPE, PP/LDPE, and PP/HDPE bilayer systems.

For both pure polymers and bilayers, the dynamic shear moduli were
measured in parallel-plate geometry with a plate diameter of 25~mm in a
rheometer~(Rheometric Dynamic Analyzer II~, TA instruments) at different
temperatures under a nitrogen atmosphere.
The measurement was performed within the frequency range
0.1-\(5\times 10^{2}\) rad/s.
The strain amplitude was set at approximately 10\% or lower, which corresponded to the
linear response regime in all of the pure polymers and bilayers, as
identified through preliminary amplitude sweeps.
Bilayer samples were prepared in the sample chamber of the rheometer.
We loaded two different polymer plaques in the sample chamber and raised
the chamber temperature to a given measurement temperature to melt the polymers.
At the measurement temperature, the gap between the parallel plates was
compressed to approximately 2~mm.
We note that the thicknesses of two polymers in a bilayer at a given measurement
temperature differ because of the difference in each polymer's thermal expansion.
To equilibrate the polymer-polymer interface, we held the bilayer at
the measurement temperatures for 30~min.
Subsequently, dynamic measurements were obtained.

Thickness fractions for each bilayer at a given measurement temperature were
computed from a digital photograph of the sample chamber up to two
decimal places.
\section{Results and Discussion}

\subsection{Linear dynamic moduli and chain architectures of the pure polymers}

\begin{figure}[htbp]
\center
\includegraphics[width=1.25\hsize]{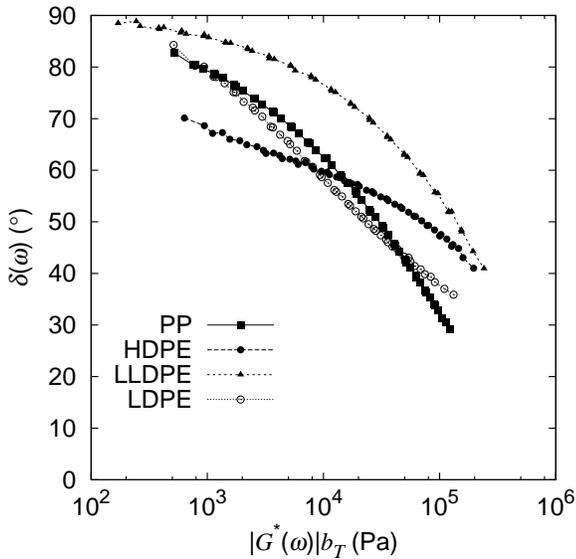}
 \caption{The van Gurp--Palmen plots of the pure polymers at a reference
 temperature of 190\degree.
}
 \label{fig:vgp190monolayer}
\end{figure}

Frequency-sweep measurements were made for each polymer listed
in Table~\ref{tbl:samples} at different temperatures between \(T_{m}\)
and \(T_{d}\):
180, 190, and 220\degree for PP 
and 140, 190, and 220\degree for LLDPE, LDPE, and HDPE, respectively.
Figure~\ref{fig:vgp190monolayer} shows the linear dynamic modulus of 
each polymer in the form of a van Gurp--Palmen plot~(phase lag versus
absolute value of the complex modulus).
In the van Gurp--Palmen plot, the frequency is eliminated  to examine the applicability of time-temperature superposition.
Figure~\ref{fig:vgp190monolayer} shows that for all the pure polymers, a
master curve was obtained by applying both frequency shift and amplitude
shift.

Next, we consider the chain architectures of the pure polymers.
It was reported that the van Gurp--Palmen curves had 
shapes specific to the different chain architectures
\cite{dealy06:_struc_and_rheol_of_molten_polym,Lohse2002WellDefined,Trinkle2002Van,schlatter05:_fourier_trans_rheol_branc_polyet,VictorHugoRolonGarrido2007Molecular,Malmberg2002LongChain}.
Although the frequency range was limited in our measurements, 
the van Gurp--Palmen curves of the pure polymers in Fig. \ref{fig:vgp190monolayer}
exhibited different
characteristic shapes.
For PP and LLDPE, the curves started from nearly
\(\delta=90^\circ\) and exhibited cap-convex shapes, 
which are characteristic of linear polymers.
For LDPE, the curves also started from nearly 
\(\delta=90^\circ\) but decreased more rapidly than a linear polymer and had
a cup-convex shape, 
which is characteristic of the long-chain-branching of densely branched polymers.
For HDPE, the curves started from \(\delta\approx 70^{\circ}\) and had a
lower \(\delta\) than LLDPE, which manifested a higher elasticity than LLDPE.
The difference in the van Gurp--Palmen curves of HDPE and LLDPE might be
explained by the difference in polydispersity and/or a possible difference in
chain architectures.
The higher the degree of polydispersity, the more the van Gurp--Palmen curve
is stretched at low modulus.\cite{Trinkle2001Van}
The large value of \(M_{w}/M_{n}=13.3\) for HDPE seems to explain in part the van
Gurp--Palmen curve for HDPE. 
However, the smaller \(\delta\) in the low-modulus region is still far from the
curves of linear chains.
This characteristic was similar to the curves of the long-chain-branching
of sparsely branched polymers~(\(<\)1 branch per chain).
\cite{Trinkle2002Van,schlatter05:_fourier_trans_rheol_branc_polyet}
To summarize these observations, the three polyethylenes used in this study
were inferred to have different chain architectures: linear~(LLDPE),
densely branched~(LDPE), and sparsely branched~(HDPE) architectures.

\subsection{Linear dynamic moduli of bilayers and non-stick degrees of interfaces}
\begin{figure}[htbp]
\center
 \includegraphics[width=1.25\hsize]{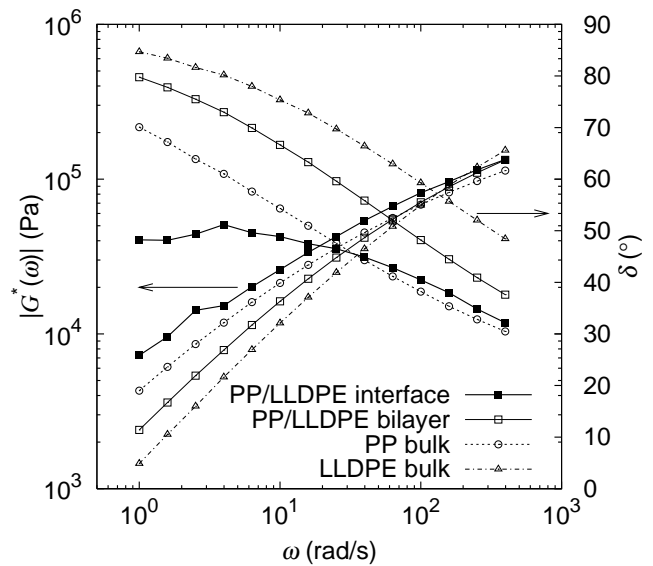}
\caption{Amplitude of the complex modulus and phase lag for PP, LLDPE,
 PP/LLDPE bilayer, and PP/LLDPE interface at 190\degree.
}
\label{fig:pp_lldpe_g}
\end{figure}

\begin{figure}[htbp]
\center
 \includegraphics[width=1.25\hsize]{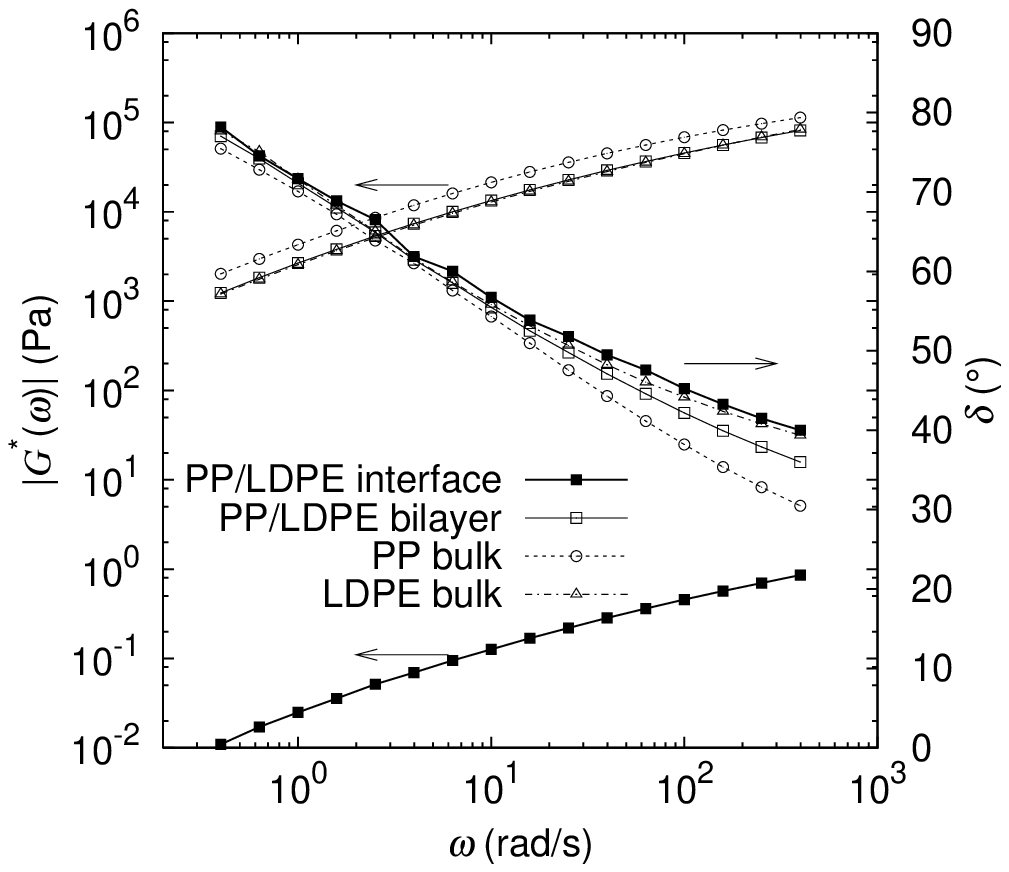}
\caption{Amplitude of the complex modulus and phase lag for PP, LDPE,
 PP/LDPE bilayer, and PP/LDPE interface at 190\degree.
}
\label{fig:pp_ldpe_g}
\end{figure}

\begin{figure}[htbp]
\center
 \includegraphics[width=1.25\hsize]{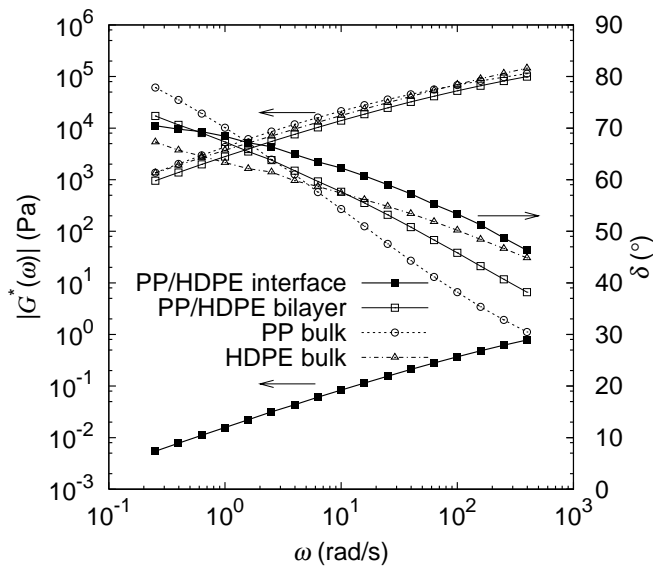}
\caption{Amplitude of the complex modulus and phase lag for PP, HDPE,
 PP/HDPE bilayer, and PP/HDPE interface at 190\degree.
}
\label{fig:pp_hdpe_g}
\end{figure}

The linear moduli of polypropylene and polyethylene bilayers are
depicted in Figs.~\ref{fig:pp_lldpe_g}, \ref{fig:pp_ldpe_g}, and
\ref{fig:pp_hdpe_g}.
The modulus amplitude and phase lag of the PP/LLDPE bilayer lies
between those of pure PP and LLDPE
~(Fig. \ref{fig:pp_lldpe_g}).
At a glance, the dynamic modulus of the PP/LLDPE bilayer appears to be an
average of those of the component polymers.
However, even for this case, some interface contribution was detected, as
described below.
Unlike PP/LLDPE, the bilayer modulus of PP/LDPE was
close to that of LDPE~(Fig. \ref{fig:pp_ldpe_g}).
This result implies that the PP/LDPE bilayer was softer than the average of
the component polymers due to an interface effect.
More distinct softening was observed in PP/HDPE
bilayer~(Fig. \ref{fig:pp_hdpe_g}).  
The bilayer modulus of PP/HDPE was lower than those of the component
polymers.
Moreover, a larger phase lag of the PP/HDPE bilayer than those of the component
polymers occurred in a certain frequency range.
This observation in the PP/HDPE bilayer also implies an interface effect. 

We now evaluate the non-stick degrees of the interfaces between
polypropylene and different polyethylenes.
For the measured moduli of a bilayer and its two component polymers at a given
frequency, \((c_{A}^{\text{st}}, c_{B}^{\text{st}})\) are obtained by
solving Eq. (\ref{eq:interface_modulus_stick}).
By definition, \(c_{I}^{\text{ns}}\) might depend on frequency.
However, the sign of \(c_{I}^{\text{ns}}\) would be insensitive to 
frequency because the softness or hardness of an interface is qualitatively
determined by the pair of polymers.
Therefore, we focus on one representative \(c_{I}^{ns}\) for each pair.
From a technical viewpoint, we computed \(c_{I}^{ns}\) at a frequency
at which the condition number of the matrix in Eq. (\ref{eq:stick_modulus})
is minimum and where \(c_{I}^{ns}\) is least affected by the measurement
error of \(G_{I}^{*}\).

The non-stick degrees \(c_{I}^{\text{ns}}\) were 
 -0.023 for the PP/LLDPE 
interface, 
0.17 for the PP/LDPE interface, and 0.32 for the PP/HDPE interface.
This result revealed that each interface contributed to the dynamic
response of each bilayer at a certain level in all the pairs of
the polypropylene and polyethylenes.
Depending on the sign of \(c_{I}^{\text{ns}}\), the interfaces are
classified into two groups.
The positive \(c_{I}^{\text{ns}}\) for PP/LDPE and PP/HDPE 
showed that these bilayers had a softer dynamic response than the averages of
the components.
In contrast, 
the negative \(c_{I}^{\text{ns}}\) for PP/LLDPE
 indicated a
hardening contribution of the interface to the dynamic response of the
bilayer.
The three polyethylenes used were similar in weight-averaged molecular
weights but differed in molecular weight distribution and chain
architecture.
The difference in the responses of the interfaces indicated that the dynamical structure
of an interfacial boundary layer depended on the chain structure rather
than solely on the miscibility.
It is presumed that a substantial chain component comprising an interface
differs depending on the chain structures.

In the PP/LLDPE bilayer, both PP and LLDPE are entangled linear chains.
In this case, the chains comprising the interface can have a number of
entanglement points.
This entanglement would make the PP/LLDPE interfacial boundary layer hard.
In contrast, HDPE has a larger polydispersity than LLDPE, indicating
that HDPE has a certain proportion of shorter chains.
Short chains are more likely to be near the interface than in the bulk.
Therefore, at the PP/HDPE interface, the number of entanglement points would
be less than at the PP/LLDPE interface, which would make the PP/HDPE interface
boundary layer soft.

In the case of a long-chain-branching LDPE, the PP/LDPE interface also
showed a softening behavior.
This fact indicates that short side chains mainly contributed to the
interface structure.
Densely branched chains are less likely to diffuse into the interfacial
layer.
Thus, backbone chains would tend to be in the bulk.
This feature could be a possible cause of a softening contribution of the
PP/LDPE interface.
The \(c_{I}^{\text{ns}}\) for the PP/LDPE interface was smaller than that
for the PP/HDPE interface.
This difference would be due to the lower mobility of a densely branched chain.

These results suggest that chain lengths and chain architectures are
responsible for the dynamic response of an interface.
Equilibrium miscibility is not sufficient to predict the dynamic response of an interface.

\subsection{Dynamic modulus of interface}
We now consider the dynamic moduli of interfaces between the polypropylene
and different polyethylenes.
To compute \(G_{I}^{*}\) by Eqs. (\ref{eq:interface_modulus_ds_dl}) and
(\ref{eq:interface_tandelta_ds_dl}), the thickness fractions of the bulk
layers A and B and the interfacial boundary layer are required.
The thicknesses of the layers A and B at a given measurement temperature were
estimated from the digital image of a bilayer sample in the chamber.
However, the thickness of an interfacial boundary layer \(h_{I}\) is not
known a priori.
For the softening interfaces of PP/LDPE and PP/HDPE,
we assume that \(h_{I}\) is of the same order of magnitude as the equilibrium
thickness of an interface \(a_{I}\), which is a lower bound of the
thickness of the interfacial boundary layer.
For a hardening interface of PP/LLDPE, the non-stick degree indicates that
the thickness of the interfacial boundary layer is much larger than the
equilibrium thickness.
There is no such large length scale in equilibrium.
Therefore, we assume an arbitrary \(h_{I}\) for the PP/LLDPE interface to
estimate its dynamic modulus, 
at least
qualitatively, 
in a physically consistent manner.

In mean-field theory, the thickness of an interface is~\cite{Jones1999Polymers,helfand71:_theor}
\begin{align}
 a_{I}&= 2\sqrt{\frac{b_{A}b_{B}}{6\chi}},
\label{eq:helfand_tagami}
\end{align}
where \(b_{\alpha}\)(\(\alpha=A,B\)) and \(\chi\) are the Kuhn statistical
segment length of the polymer \(\alpha\) and the Flory--Huggins interaction
parameter between polymers A and B,
respectively.
The Kuhn lengths have been measured or may be estimated for many
polymers.\cite{fetters94:_connec_polym_molec_weigh_densit,fetters99:_chain,mark06:_physic_proper_polym_handb}
Flory--Huggins parameters are estimated
by~\cite{brandrup99:_polym_handb_edition,mark06:_physic_proper_polym_handb,young11:_introd_polym_third_edition}
\begin{align}
 \chi&=\frac{V_{ref}}{RT}\left(\delta_{1}-\delta_{2}\right)^{2},
\label{eq:chi_regular_solution}
\end{align}
where \(V_{ref}\) is an equivalent monomer reference volume, which might be the geometric mean of two components, \(\delta_{\alpha}\) is the solubility parameter of the polymer \(\alpha\), \(R\) is the gas constant, and \(T\) is the temperature.
Using Eqs. (\ref{eq:helfand_tagami}) and (\ref{eq:chi_regular_solution}), the equilibrium interface thickness between polypropylene and polyethylene
is estimated to be \(a_{I}=3.58\) nm.
In addition to the mean-field prediction, the thermal capillary wave correction
is required to improve the accuracy compared to experimental
measurements of the equilibrium interface thickness.\cite{Jones1999Polymers}
However, this correction does not change the order of the prediction.
Thus, we neglect the thermal capillary wave correction in \(a_{I}\).
The effect of long-chain-branching on the Flory--Huggins parameter is
not accounted for in Eq. (\ref{eq:chi_regular_solution}) but should not change the order of \(a_{I}\).
From the digital image of a bilayer sample, the apparent thickness fractions
of the bulk layers, \(c'_{A}\) and \(c'_{B}\), were measured, where
\(c'_{A}+c'_{B}=1\).
Combining an assumed \(h_{I}\) with the apparent thickness fractions,
the thickness fractions are estimated as
\(c_{\alpha}=c'_{\alpha}-c_{I}/2\)~(\(\alpha=A,B\)) where
\(c_{A}+c_{B}+c_{I}=1\) holds.
The interface moduli, \(G_{I}^{*}\), of PP/LLDPE, PP/LDPE, and PP/HDPE
estimated using Eqs. (\ref{eq:interface_modulus_ds_dl}) and
(\ref{eq:interface_tandelta_ds_dl}) are shown in
Figs. \ref{fig:pp_lldpe_g},~\ref{fig:pp_ldpe_g}, and \ref{fig:pp_hdpe_g},
respectively.
In the case of PP/LLDPE, \(h_{W}\left|c_{I}^{\text{ns}}\right|\approx 48
\mu\)m, which gives a lower bound of the thickness of the interfacial
boundary layer.
For the strain of the interfacial boundary layer,
\(c_{I}\gamma_{I}=\gamma_{W} -c_{A}\gamma_{A} -c_{B}\gamma_{B}\),
to be positive,
a larger length scale than
\(h_{W}\left|c_{I}^{\text{ns}}\right|\) should exist.
A value of at least \(240~\mu\)m for \(h_{I}\) was required for the interface phase
lag to be positive.
This fact indicates that the thickness of the interfacial boundary layer
for PP/LLDPE is much greater than the gyration radii of the
components.
This length scale is considered to be associated with the collective
motion of chains under dynamic shear.
In Fig. \ref{fig:pp_lldpe_g}, \(G_{I}^{*}\) is shown for the PP/LLDPE interface estimated with
\(h_{I}=300~\mu\)m.
The amplitude \(\left|G_{I}^{*}\right|\) for PP/LLDPE was larger than
those of the bulk layers and the bilayer when \(h_{I}\) was assumed to be in the range of
240-400\(\mu\)m.
Moreover, the interface phase lag was estimated to be lower than those
of the bulk layers and the bilayer at a low frequency, indicating that the
interfacial boundary layer of PP/LLDPE had an additional slower
relaxation mode compared to the bulk layers.
Concerning softening interfaces, \(G_{I}^{*}\) for the PP/LDPE and PP/HDPE
interfaces estimated with \(h_{I}=a_{I}\) are shown in
Figs.~\ref{fig:pp_ldpe_g} and \ref{fig:pp_hdpe_g}, respectively.
In these cases, the amplitudes of the interface dynamic moduli were 
much lower than those of bulk layers.
Moreover, the interface phase lags were larger than those of the bulk
layers and the bilayer, indicating that relaxations in the interfacial
boundary layers of PP/LDPE and PP/HDPE were faster than in the bulk layers.
These observations were irrespective of the choice of \(h_{I}\),
although the absolute values of \(\left|G_{I}^{*}\right|\) and
\(\tan\delta_{I}\) depend on the scale \(h_{I}\).

The results revealed that the dynamic shear response of a polymer-polymer
interface is different from those of the bulk components in both the
amplitude and the phase lag, indicating the existence of an
interfacial boundary layer, which is a finite region with a different
dynamical response than the bulk components.
The difference in the phase lag indicates the existence of additional
relaxation modes in the interfacial boundary layer.
Moreover, the characteristic thickness scale of the interfacial boundary
layer differed from the equilibrium interface thickness predicted
by mean-field theory.
This result would indicate that both polymer chains within the
equilibrium interface and those in the bulk region near the
interface collectively contribute to a relaxation of the interfacial
boundary layer.
Therefore, the length and architecture of the chains near the interface
are supposed to strongly affect the dynamical response of the
interfacial boundary layer.
These results support this view.
\section{Concluding Remarks}
In this article, we developed a way to assess the rheological
contribution of an interface to a bilayer system under small-amplitude
oscillatory shear.
A non-stick degree of interface was proposed to quantify the deviation
from stick boundary conditions.
The application of the method to three immiscible bilayer systems of 
polypropylene and different polyethylenes revealed that
the dynamic shear response of the interfacial boundary layer depended on the
chain length, polydispersity, and architecture of the component polymers
rather than the equilibrium miscibility.

The interface between two linear, long polymers showed a hardening
contribution to the bilayer, which was characterized by a negative
non-stick degree or negative slip length.
The thickness of the hardening interfacial boundary layer would be much larger
than the equilibrium thickness determined by mean-field theory,
indicating the existence of collective motion of chains near the
interface.
The interface between linear, long polypropylene and long-chain-branching
polyethylene 
and the interface between linear, long polypropylene and a polyethylene
with large polydispersity showed a softening contribution to the bilayer,
which was characterized by a positive non-stick degree or positive slip
length.
Based on the estimation of the dynamic moduli of the interfaces, the phase lag of an
interfacial boundary layer was different from that of the bulk layers.
The interfacial boundary layer was more elastic in the hardening case and
more viscous in the softening case.
These findings suggest that interfacial boundary layer has different
relaxation modes than the bulk layers.
Further studies on chain dynamics around the interface are required to
determine the length scales and structure of the interfacial boundary layer.

The proposed method can be applied to other systems having complex interfaces
with internal structures.
For soft bulk materials,
the dynamic response is a convenient and useful probe to study
the internal relaxation structure.
Therefore, the proposed method is a useful rheological tool for
investigating the dynamic structure of complex interfaces, including
copolymer compatibilized interfaces, colloid/nanoparticle-absorbed
interfaces, and biological interfaces.

%


\end{document}